\documentclass[conf]{new-aiaa}

\usepackage[utf8]{inputenc}
\usepackage[version=4]{mhchem}
\usepackage{siunitx}
\usepackage{longtable,tabularx}
\usepackage{amsmath}
\usepackage{algorithmic}
\usepackage{graphicx}
\usepackage{textcomp}
\usepackage{xcolor}
\usepackage{wrapfig}
\usepackage{tikz}
\usetikzlibrary{quantikz}

\newcommand{\afigref}[1]{\text{Fig.}~\ref{#1}} 

\newcommand{\avg}[1]{{\left< {#1}\right> }}

\usepackage{stfloats}

\usepackage{comment}
\usepackage{ulem} 
\usepackage{braket}

\newcommand\eq[1]{Eq.~(\ref{eq:#1})}
\newcommand\Eq[1]{Equation~(\ref{eq:#1})}

\newcommand\fig[1]{Fig.~\ref{fig:#1}}
\newcommand\om[1]{\omega_{\scriptscriptstyle #1}}

\graphicspath{./figures/}

\title{Implementing the Koopman-von Neumann approach on continuous-variable photonic quantum computers}

 \author{Xinfeng Gao\footnote{Professor, Department of Mechanical and Aerospace Engineering, x.gao@virginia.edu, and AIAA Associate Fellow.}, Olivier Pfister\footnote{Professor, Department of Physics and Charles L. Brown Department of Electrical and Computer Engineering, opfister@virginia.edu}, \ Stefan Bekiranov\footnote{Professor, Department of Biochemistry and Molecular Genetics, sb3de@virginia.edu}}
 \affil{University of Virginia, Charlottesville, VA, 22904, USA}

\begin{document}

\maketitle

\begin{abstract}
The Koopman-von Neumann (KvN) formalism recasts classical mechanics in a Hilbert space framework using complex wavefunctions and linear operators, akin to quantum mechanics. Instead of evolving probability densities in phase space (as in Liouville’s equation), KvN uses a Schr\"{o}dinger-like equation for a classical wavefunction, with commuting position and momentum operators. Mapped to quantum computing, KvN offers a promising route to simulate classical dynamical systems using quantum algorithms by leveraging unitary evolution and quantum linear algebra tools, potentially enabling efficient classical-to-quantum mappings without invoking full quantum uncertainty. In this work, we specifically explore the implementation of the KvN approach on continuous-variable photonic quantum computing architectures, with the goals of leveraging quantum simulation for both sampling and computing intractable nonlinear dynamics. We will demonstrate its implementation and feasibility with two problems: the harmonic oscillator and a 1D partial differential equation governing nonlinear dynamics.

\end{abstract}

\section{Introduction}

\lettrine{T}{he} Koopman von Neumann (KvN) formalism offers the prospect of quantum advantage for quantum simulation of classical nonlinear dynamics. The KvN formalism retains the classical phase space, that is, the use of classical variables—position $x$ and momentum $p$ and the phase space structure of classical mechanics, but uses wavefunctions and operators to describe dynamics in a linear, Hilbert space setting. In doing so, KvN enables the use of spectral theory and functional analysis tools traditionally reserved for quantum systems, allowing for deeper insights into classical dynamics. Moreover, it provides a unified language to bridge classical and quantum mechanics, facilitating studies of quantum-classical correspondence, decoherence, and the emergence of classicality. KvN is also particularly valuable in the analysis of long-term behavior and chaos, where Koopman operators serve as a rigorous tool for studying ergodic properties and the spectral characteristics of classical systems.

In this paper, we explore mapping the KvN formula onto continuous-variable quantum computing (CVQC) platforms. Our overarching goal is to develop practical quantum algorithms that are capable of simulating nonlinear nonequilibrium gas kinetics. As an intermediate goal, we will investigate realizable and efficient approaches for mapping the continuous solution variables using qumodes onto a quantum computer. Specifically, we will demonstrate our approach with the Korteweg–De Vries (KdV) equation, describing 1D nonlinear dispersive nondissipative waves.

\section{Background and Methodology}

\subsection{Derivation of the KvN Formalism}

The KvN formalism was initially formulated in terms of classical Liouvillian mechanics but can be cast in a more general form~\cite{Joseph2020,Barthe2023,Cochran2025} that is of direct interest to our approach of using CVQC to simulate classical dynamics. Consider the following, arbitrary ODE
\begin{align}\label{eq:ode}
\dot{\mathbf u} \coloneqq \frac{d\mathbf u}{dt} = \mathbf{v}(\mathbf{u}),
\end{align}
where $\mathbf u(t)\in\mathbb R^{2n}$ are phase space variables with $n$ being the degrees of freedom and $\mathbf{v}$ is an arbitrary vector field~\cite{Joseph2020} which can be generally thought of as nonlinear functions of $\mathbf u$. Notably, \eq{ode} includes Hamilton's equations of motion as well as non-Hamiltonian dynamics such as the KdV equation. Importantly, $\mathbf u$ can represent any physical or mathematical dynamical variables as long as their dynamics is described by \eq{ode}.  We provide an approach to simulating both classes of classical dynamic equations. Similar to Cochran et al.~\cite{Cochran2025} in the case of Hamiltonian dynamics, we will define $u_{1}, u_{2},\dots,u_{n}$ to be the generalized position coordinates, which are usually defined as $q_{1}, q_{2},\dots,q_{n}$, and conjugate momenta as $u_{n+1}, u_{n+2},\dots,u_{2n}$, which are usually defined as $p_{1}, p_{2},\dots,p_{n}$, for a $2n$-dimensional phase space. Using this definition of the phase space coordinates, Hamilton's equations of motion are    
\begin{align}
    \dot{u}_{i} &= \frac{\partial H(\mathbf{u}(t),t)}{\partial u_{n+i}} \label{eq:Ham1} \\
    \dot{u}_{n+i} &= -\frac{\partial H(\mathbf{u}(t),t)}{\partial u_{i}} \label{eq:Ham2},
\end{align}
for $i=1,2,\dots,n$. Eqs.~\eqref{eq:Ham1} and \eqref{eq:Ham2} define the relationship between velocity and momentum and Newton's second law, respectively. Defining the phase space variables in this way, Eqs.~\eqref{eq:Ham1} and \eqref{eq:Ham2} clearly take the form of \eq{ode}. Note that we will define a momentum operator $\mathbf{P}$ and detail it below.

Without loss of generality, one can define a probability density in phase space $\rho(\mathbf{u},t)$ which satisfies the continuity equation
\begin{equation} \label{eq:cont}
    \frac{\partial \rho(\mathbf{u},t)}{\partial t} + \nabla \cdot (\mathbf{v} \rho(\mathbf{u},t)) = 0, 
\end{equation}
where $\mathbf{v}$ is the vector field associated with a conserved quantity (e.g., mass in the case of fluid dynamics or phase space density in the case of Hamiltonian dynamics), and the expectation of any observable $O(\mathbf{u},t)$, including the phase space vector, $\mathbf{u}$, is
\begin{equation} \label{eq:exp}
    \langle O(\mathbf{u},t) \rangle = \int O \rho(\mathbf{u},t)d\mathbf{u}.
\end{equation}

The first step in deriving the KvN dynamical equations is to define a KvN wavefunction or probability amplitude, $\psi(\mathbf{u},t)$, in the position representation in terms of the probability density, just as in the case of quantum mechanics~\cite{Joseph2020}
\begin{equation} \label{eq:kvnrho}
    \rho(\mathbf{u},t) = \psi(\mathbf{u},t) \psi^{*}(\mathbf{u},t). 
\end{equation}

Substituting \eq{kvnrho} into \eq{cont} and applying the time and spatial derivatives enables the separation of the result into two self-consistent equations
\begin{equation} \label{eq:psiinter}
    \frac{\partial \psi(\mathbf{u},t)}{\partial t} = - \Bigl ( \frac{1}{2} \psi(\mathbf{u},t) \nabla \cdot \mathbf{v} + (\mathbf{v} \cdot \nabla) \psi(\mathbf{u},t) \Bigr ),    
\end{equation}
and the complex conjugate of \eq{psiinter}. We can rewrite this equation in the following Schr\"{o}dinger equation-like form by allowing the first del operator to act on $\psi(\mathbf{u},t)$ as well as $\mathbf{v}$ and multiplying both sides by $i \hbar$
\begin{equation} \label{eq:KvNDel}
    i \hbar \frac{\partial \psi(\mathbf{u},t)}{\partial t} = \frac{-i \hbar}{2} \Bigl (\nabla \cdot \mathbf{v} + \mathbf{v} \cdot \nabla \Bigr ) \psi(\mathbf{u},t),  
\end{equation}
where we note for clarity that the del operator in the first term on the right acts on both $\mathbf{v}$ and $\psi(\mathbf{u},t)$. The second major step in deriving the KvN formalism is to define a momentum operator $\mathbf{P} = -i \hbar \nabla$ and a position operator $\mathbf{Q} = \mathbf{u}$. Thus, using these operator definitions, we arrive at the KvN equation which takes a Schr\"{o}dinger equation form~\cite{Joseph2020}
\begin{equation} \label{eq:KvNSchro}
    i \hbar \frac{\partial \psi(\mathbf{Q},t)}{\partial t} = H_\text{KvN} \psi(\mathbf{Q},t),
\end{equation}
where
\begin{gather}
 {H_{\text{KvN}}}=\frac{1}{2} \Bigl [{\textbf{P}}\cdot\textbf{v}\left({\textbf{Q}}\right)+\textbf{v}\left({\textbf{Q}}\right)\cdot{\textbf{P}} \Bigr ] \label{eq:HKvN}
\end{gather}
is the KvN Hamiltonian operator which is Hermitian. Given that the momentum and position operators are defined in the same way as they are in the standard formulation of quantum mechanics, they satisfy the well known position-momentum commutation relations
\begin{equation} \label{eq:QPCommute}
    [Q_{i},P_{j}] = i \hbar \delta_{ij},    
\end{equation}
where $\delta_{ij}$ is the Kronecker delta function (i.e., 1 if $i=j$ and 0 otherwise). Finally, the time evolution of the KvN wavefunction is given by a solution of \eq{KvNSchro} in ket form which yields the well known unitary time evolution operator
\begin{equation} \label{eq:KvNUnitary}
    \ket{\psi(t)} = e^{-i H_\text{KvN} t/\hbar} \ket{\psi(0)}, 
\end{equation}
from which the Heisenberg equation of motion may be derived
\begin{align}\label{eq:KvN}
    \dot{\mathbf Q} = \frac1{i\hbar} \Bigl [ \mathbf Q,H_\text{KvN} \Bigr ] = \mathbf{v}(\mathbf{Q}),
\end{align}
where we have used Eqs.~\eqref{eq:HKvN} and \eqref{eq:QPCommute} to arrive at the final equality. We note that this demonstrates that the KvN formalism is self consistent in that \eq{KvN} is the original classical equation of motion \eq{ode}~\cite{Joseph2020}. In summary, this equation describes the evolution, under the KvN Hamiltonian, of a $2n$-partite quantum system simulating the classical system described by \eq{ode}.

In the particular case of quantum optics, $Q_j=(a_j+a^\dag_j)/\sqrt2$ and $P_j=i(a^\dag_j-a_j)/\sqrt2$ are the amplitude and phase quadratures of qumode $j$ of the quantum optical field, $a_j$ and $a_j^\dag$ being the photon annihilation and creation operators for qumode $j$. We have the canonical commutation relations $[Q_j,P_k]=i\delta_{jk}$, yielding the Heisenberg inequalities $\Delta Q_j\Delta P_j\geqslant 1/2$.

In the next section, we show how to leverage quantum optics technology to propose concrete photonic implementations of KvN Hamiltonians.

\subsection{Quantum Algorithm}

A practical quantum simulator implementing a KvN Hamiltonian can then be employed to quantum sample \eq{ode}, starting from a well-defined initial state $\mathbf u(t=0)=\mathbf u_o$. Running the quantum KvN circuit and measuring $\mathbf{Q}$ at different times $t$ yields two applications:
\begin{enumerate}
    \item The generation of simulated classical data by quantum sampling of $\mathbf Q$ over a probability distribution of $\mathbf u$.
    \item The computation of $\mathbf u(t)=\langle \mathbf Q(t)\rangle$~\cite{Barthe2023}.
\end{enumerate}

\section{KvN numerical case studies and proposed CVQC experiments}

We will consider several test problems: the simple harmonic oscillator, its extension to many coupled oscillators, leveraging the inherent scalability of quantum optics, and the more challenging nonlinear KdV equation.

\subsection{Single harmonic oscillator}

We first consider an extremely simple system in order to showcase the KvN formalism and a fully fledged quantum photonics implementation for simulating nanomechanics.

\subsubsection{General KvN formalism}
Using the variables that define the generalized coordinate and conjugate momentum variables as shown in Eqs.~\eqref{eq:Ham1} and \eqref{eq:Ham2}, the Hamiltonian of the simple classical harmonic oscillator (HO)
\begin{equation}\label{eq:ham}
    H = \frac{1}{2m}  {u_{2}}^2 + \frac{1}{2} m \om o^2  {u_{1}}^2, 
\end{equation}
where $u_{1}$ and $u_{2}$ denote the classical position and momentum, respectively, and, using Eqs.~\eqref{eq:Ham1} and \eqref{eq:Ham2}, the equations of motion are
\begin{align}
    \dot u_{1} &= \frac1m u_{2}\\
    \dot u_{2} &= -m\om o^2 u_{1}.
\end{align}
The KvN approach to this toy problem consists in assigning a quantum mode (qumode) $Q_{1,2}$ for each of the classical variables $u_{1,2}$. We therefore seek Heisenberg equations of the form
\begin{align}
    \dot Q_1 &= \frac1m Q_2\\
    \dot Q_2 &= -m\om o^2 Q_1,
\end{align}
which are obtained from the {\it two-mode} KvN Hamiltonian for a single oscillator, \eq{HKvN}, and the commutation relation, \eq{QPCommute}, 
\begin{align}\label{eq:KvNSO}
 H_\text{SO}  = \frac{1}{m} P_1Q_2  - m \omega^2 Q_1 P_2, 
\end{align}
where $Q_j$ and $P_j$ are now {\it quantum} position and momentum operators.
Note that, while replacing $u_{1}$ with $Q_1$ poses no problem, replacing $u_{2}$ with $Q_2$ does entail a dimension change and dimensional analysis of \eq{KvNSO} shows that $m$ has become homogeneous to a time, while $\om o$ remains homogeneous to a frequency. In simulation practice, the numerical values of these parameters $m$ and $\om o$ will not change from the initial physical ones, only the units may change. We will note another important unit change in a moment.

As was proposed in Ref.~\cite{Cochran2025}, this KvN-HO Hamiltonian can be formally implemented by a Trotter-Suzuki expansion of two controlled-X gates [$CX_{12}(\alpha) := \exp(-i\alpha Q_1P_2)$],
\begin{align}\label{eq:HO}
 e^{-i\tfrac\tau\hbar H_\text{SO}}\simeq \left[e^{-i\frac\tau\hbar\frac{1}{pm} P_1 Q_2}\,e^{i\frac\tau\hbar\frac{m\omega^2}p  Q_1 P_2} \right]^p.  
\end{align}
We now propose a new, easier method for implementing this by use of quantum optics. 

\subsubsection{Photonic CVQC KvN implementation}

Gaussian (Wigner function) quantum optics readily provides usable quadratic Hamiltonians, namely the beam splitter (BS) 
\begin{align}\label{eq:bs}
H_\text{BS}&= i\hbar \alpha (a_1a^\dag_2-a^\dag_1a_2)\\
&= \hbar \alpha (P_1Q_2-Q_1P_2),
\end{align}
e.g.\ a partially reflecting mirror, and the two-mode squeezer (TMS)
\begin{align}
H_\text{TMS}&= i\hbar \beta (a^\dag_1a^\dag_2-a_1a_2)\\
&= \hbar \beta (P_1Q_2+Q_1P_2),
\end{align}
e.g.\ an optical parametric amplifier based on stimulated (rather than spontaneous) photon pair emission, say in a nonlinear optical material.
Using the identity
\begin{align}
    \gamma P_1Q_2+\delta Q_1P_2 = 
    \tfrac{\gamma +\delta}{2}(P_1Q_2+Q_1P_2)
    +\tfrac{\gamma -\delta}{2}(P_1Q_2-Q_1P_2),
\end{align}
we rewrite $H_\text{SO}$ in \eq{KvNSO} as
\begin{align}\label{eq:TrotterPhotonics!}
 H_\text{SO}= \frac\hbar2(\frac{1}{m}-m \omega^2)(P_1Q_2+Q_1P_2)+\frac\hbar2(\frac{1}{m}+m \omega^2) (P_1Q_2-Q_1P_2), 
\end{align}
where we have renormalized the canonical operators $Q_j$, $P_j$ to quantum optical fields, namely:
\begin{align}
    Q &\mapsto \frac{Q}{\beta},\\
    P &\mapsto \hbar\beta P,
\end{align}
with $\beta=(m\om{}/\hbar)^{1/2}$. (Note that $\beta$ plays no role in $H_{\rm{SO}}$ as $PQ\mapsto\hbar PQ$). This renormalization of operators lets us define the photon annihilation operator as $a=(Q+iP)/\sqrt2$, $a^\dag a$ being the {\it dimensionless} photon number operator.

\subsubsection{Quantum optical implementation}

\begin{figure}[h]
\centerline{\includegraphics[width=.75\textwidth]{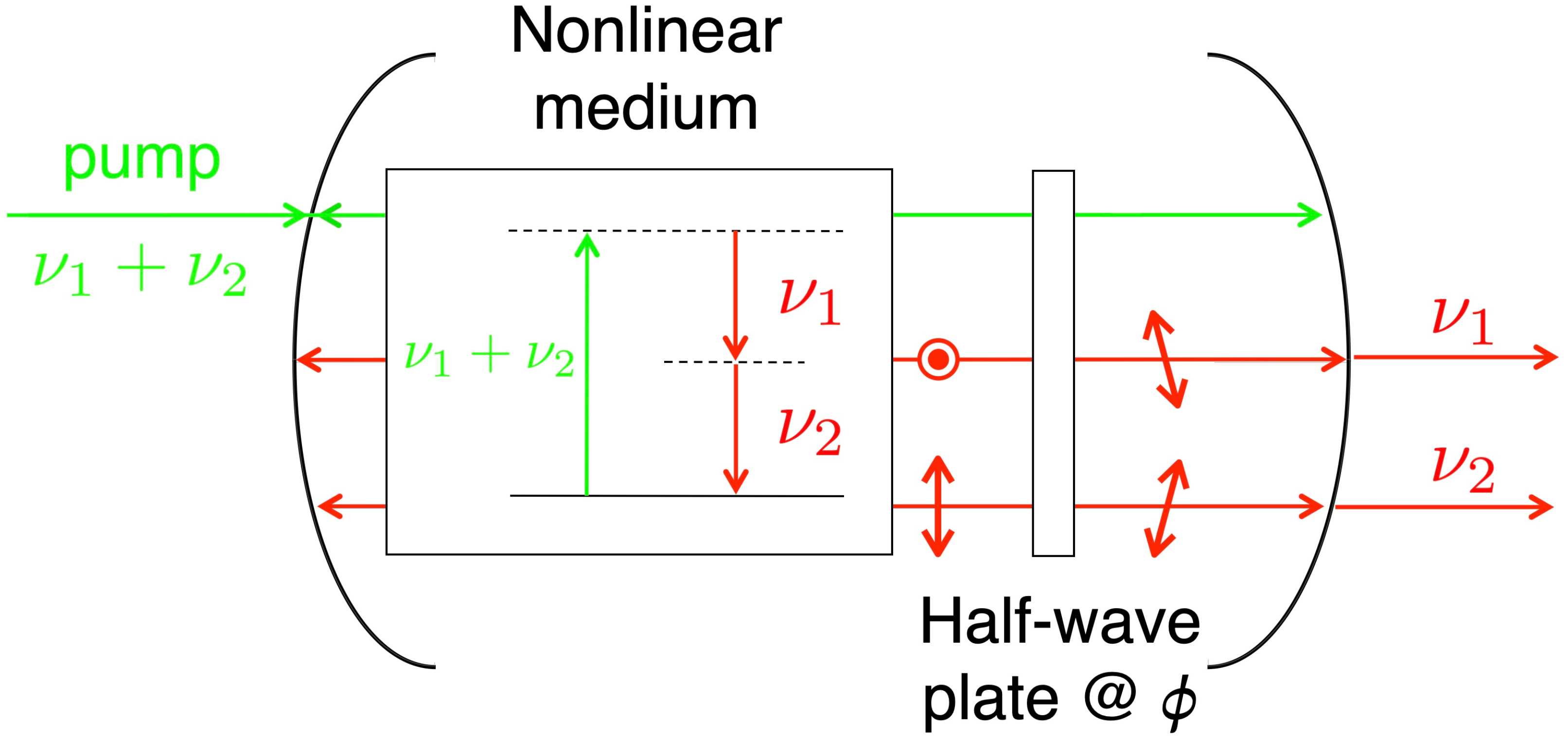}}
\caption{Principle sketch of a  Trotterizing optical parametric oscillator for the simulation of a classical harmonic oscillator. The OPO signal fields (qumodes 1 and 2 in the text) are embodied by OPO modes with orthogonal linear polarizations. The ``type-II'' OPO nonlinear medium on the left (e.g.\ KTP or PPKTP) annihilates pump photons of frequency $\nu_p=\nu_1+\nu_2$ to create pairs of orthogonally polarized signal photons at $\nu_1$ and $\nu_2$. The corresponding fields are then mixed by a half-wave plate at orientation $\phi$ equivalently to a beam splitter of angle $\theta=4\phi$ (see text). For illustration, the field polarizations are sketched qualitatively going left to right in the initial round trip . Each round trip inside the OPO cavity constitutes a Trotter-Suzuki step in \eq{TSopt}.}
\label{fig:OPO}
\end{figure}

The new form of $H_{\rm{SO}}$ in \eq{TrotterPhotonics!} lets us write the Trotter-Suzuki expansion of $H_\text{SO}$ as
\begin{align}\label{eq:TSopt}
 e^{-i\tfrac\tau\hbar H_\text{SO}}\simeq \left[e^{-i\frac\tau{2p}(\frac{1}{m}-m \omega^2)(P_1Q_2+Q_1P_2)}\,e^{-i\frac\tau{2p}(\frac{1}{m}+m \omega^2) (P_1Q_2-Q_1P_2)} \right]^p, 
\end{align}
which corresponds to the sequencing of a two-mode squeezer with a beamsplitter, respectively. Note that these two building blocks can Trotterize any Gaussian Hamiltonian. Such quantum optical Trotterization was demonstrated experimentally, to a very good approximation, by one of us (OP) on two different TMS gates in the first-ever large-scale cluster-state entanglement of an optical frequency comb~\cite{Pysher2011,PT2011}. The procedure is to place the aforementioned gates (TMS, BS) inside a resonant optical cavity. Additional compactness can be achieved by encoding the two qumodes in polarization rather than space and using a half-wave plate, as depicted in \fig{OPO}.
The device in \fig{OPO} is a type-II optical parametric oscillator (OPO), constituted by a resonant optical cavity which effectively implements the $p$ Trotter steps over $p$ round trips (on average). The OPO cavity contains a nonlinear optical crystal, such as periodically poled $\rm KTiOPO_4$ (i.e., Potassium Titanyl Phosphate), which two-mode-squeezes the two qumodes 1 and 2 by photon-pair emission. In \fig{OPO}, the qumodes are orthogonally polarized and a half-wave plate acts as a variable beam splitter. The intensity of the pump beam in the nonlinear crystal and the wave plate angle provide two independent controls for precisely adjusting the Trotter parameters $\frac{1}{m}\mp m \omega^2$ in \eq{TrotterPhotonics!}.

Let's now examine the simulation regimes that are accessible experimentally. Let us define $r$ to be the squeezing parameter of the OPO. For an OPO below the oscillation threshold, over $p$ passes, we have
\begin{align}
    Q_1-Q_2 \mapsto e^r\,(Q_1-Q_2) \\  
    P_1-P_2 \mapsto e^{-r}\,(P_1-P_2)
\end{align}
with $r=1.15$ for a 10 dB squeezing ratio. We thus have
\begin{align}
    \frac r{p\tau} = \frac12(\frac1m-m\om o^2).
\end{align}
The beam splitter field reflectivity can be expressed as $\rho = \cos\theta$ and we have 
\begin{align}
    \frac\theta\tau = \frac12(\frac1m+m\om o^2).
\end{align}
This yields the simulated oscillator parameters
\begin{align}
    m &= \tau\left(\theta+\frac rp\right)^{-1},\\
    \om o &= \frac1\tau\left(\theta^2-\frac{r^2}{p^2}\right)^\frac12.
\end{align}
Taking typical OPO parameter values $\tau= 1$ ns, $r=\pm1$, $\theta=0.1$ rad, yields $m\sim 10^{-9}$ kg and $\om o/2\pi\sim 0.1$ GHz, i.e., a nanomechanical oscillator.

\subsection{Coupled harmonic oscillator network}

We now extend this implementation to a network of coupled oscillators. This could be used in the context of a recent quantum algorithm proposed for the exponentially faster simulation of coupled oscillators~\cite{Babbush2023}.

\subsubsection{KvN Formalism}
Using the generalized coordinates and conjugate momentum variables defined in Eqs.~\eqref{eq:Ham1} and \eqref{eq:Ham2}, the classical Hamiltonian of $n$ coupled oscillators is
\begin{align}
    H_{CO}^\text{(class.)} &= \sum_{j=1}^n\left(\frac{u_{n+j}^2}{2m_j}+\frac{\kappa_j}2u_j^2\right) + \sum_{j=1}^n\sum_{k>j}^n \kappa_{jk}(u_j-u_k)^2\\
    &=\sum_{j=1}^n\left(\frac{u_{n+j}^2}{2m_j}+\frac{\xi_j}2u_j^2\right) + \sum_{j=1}^n\sum_{k\neq j}^n \xi_{jk}\, u_ju_k
\end{align}
where $\xi_{jk}=-2\kappa_{jk}=\xi_{kj}$ and $\xi_j=\kappa_j+\sum_{k\neq j}^n \kappa_{jk}$.
Following the KvN formalism, we promote $\mathbf{u} = (u_1,\dots, u_{2n})^T$ to a position operator, $\mathbf{Q} = (Q_1,\dots, Q_{2n})^T$. Using \eq{KvN}, the Heisenberg equations are
\begin{align}
    \dot{\mathbf{Q}} = 
    \begin{pmatrix}
        &&&&\frac1{m_1}&&&\\
        &&&&&\frac1{m_2}&&\\
        &&&&&&\ddots&\\
        &&&&&&&\frac1{m_n}\\
        -\xi_1&-\xi_{12}&\dots&-\xi_{1n}&&&\\
        -\xi_{12}&-\xi_2&\dots&-\xi_{2n}&&&\\
        \vdots&\vdots&\ddots&\vdots&&&\\
        -\xi_{1n}&-\xi_{2n}&\dots&-\xi_n&&&
    \end{pmatrix}\mathbf{Q},
\end{align}
where the blank entries in the matrix are zeroes. Using \eq{HKvN}, the 2$n$-mode KvN Hamiltonian is therefore, including Trotterization options:
\begin{align}\label{eq:CO}
    H_{CO} = \underbrace{\sum_{j=1}^n \frac1{m_j}P_jQ_{j+n}-\sum_{j=1}^n\xi_jP_{j+n}Q_j}_{\textcolor{green}{\text{independent oscillators (TMS+BS)}}} \underbrace{\ -\sum_{j=1}^n\sum_{k\neq j}^n \xi_{jk}P_{j+n}Q_k}_{\textcolor{blue}{\text{coupling graph (TMS only)}}}.
\end{align}

\subsubsection{Quantum photonic implementation}

As for the single harmonic oscillator, the KvN Hamiltonian can be Trotterized using beamsplitters and squeezers inside an optical cavity. Here, the platform advantage we seek is scalability: the more oscillators, the more interesting the simulation. Previous research by one of us (OP) demonstrated that one can make use of the spectral degree of freedom offered by the quantum optical frequency comb (QOFC) of resonant OPO modes as an eminently scalable platform for quantum computing~\cite{Pfister2019}. In particular, entanglement of as many as 60 qumodes into a cluster state, a quantum computing ``substrate,'' has been experimentally demonstrated~\cite{Pysher2011,Chen2014}. 

The design challenges will consist here in, {\it (i)}, employing frequency-domain beamsplitters [green term in \eq{CO}] and, {\it (ii)}, implementing the arbitrary coupling matrix, $(\xi_{jk})_{j,k}$ [blue term in \eq{CO}], needed for the simulation at hand. While addressing specific designs are out of the scope of this paper, we wish to point out that sophisticated QOFC engineering methods have already been proposed, based on optical phase modulation~\cite{Zhu2021} as well as on generation of arbitrary entanglement graphs~\cite{Wang2014a,Brunel2025}. These results will be applied to specific oscillator coupling topologies as needed.

\subsection{Korteweg-De Vries equation}

We now turn to a more arduous, nonlinear simulation problem, the Korteweg-De Vries equation, which models waves on a shallow water surface.

\subsubsection{KvN formalism}

\begin{figure}[h!]
  \centering
 \includegraphics[width=1.8in]{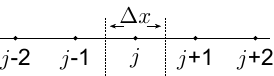}
 \caption{Control volume in 1D. \label{fig:1dgrid}}
 \end{figure}

The 1D KdV equation of a function $u (x,t)$ has the form  
\begin{equation}
\frac{\partial u}{\partial t} - 6 u \frac{\partial u}{\partial x} + \frac{\partial^3 u}{\partial x^3} = 0. \label{eq:KdV} 
\end{equation}
This problem is chosen for its dispersive and nonlinear properties, representing typical features in fluid dynamics.
In a first step, we convert the PDE of \eq{KdV} to an ODE, using a finite difference method on the 1D grid of pitch $\Delta x$ with a periodic boundary in $x$, as shown in \afigref{fig:1dgrid}, 
to yield the discretized \eq{num}. Note that the discretization scheme is not restricted to finite differences. Methods such as finite volume, finite element, or Galerkin are applicable. 
\begin{gather}
{\frac{du_j}{dt}}=v_j(u) \equiv -\frac3{\Delta x} (u_{j+1}^2-u_{j-1}^2) - \frac1{2\Delta x^3}(u_{j-2}-2u_{j-1}+2u_{j+1}-u_{j+2}). \label{eq:num}
\end{gather}
\Eq{num} will lend itself well to KvN implementation provided one has access to non-Gaussian quantum gates. From \eq{HKvN}, we obtain the corresponding KdV-KvN Hamiltonian
\begin{gather}
H_{\text{KdV}} = \underbrace{- \frac1{2\Delta x^3}\sum_{j=1}^N (P_jQ_{j-2}-2P_jQ_{j-1}+2P_jQ_{j+1}-P_jQ_{j+2})}_{\textcolor{blue}{\text{Gaussian Hamiltonian } H_G\text{ (BS only)}}} \underbrace{-\frac3{\Delta x} \sum_{j=1}^N (P_jQ_{j+1}^2-P_jQ_{j-1}^2)}_{\textcolor{purple}{\text{non-Gaussian Hamiltonian } H_{NG}}}. \label{eq:HKdV}
\end{gather}

\subsubsection{Photonic CVQC implementation}
 
As for the harmonic oscillator case, we now proceed to implement the KdV-KvN Hamiltonian on a photonic quantum machine, again using a Trotter-Suzuki expansion. 

\begin{enumerate}
\item Remarkably, close examination of the Gaussian term $\color{blue}H_G$ on the RHS of \eq{HKdV} in light of \eq{bs} reveals that it solely contains  beamsplitter terms---of field reflectivities $\cos\left(\tau/(2\Delta x^3)\right)$ and $\cos\left(\tau/(\Delta x^3)\right)$, where $\tau$ is the fixed propagation time through the beamsplitter---and can therefore simply be viewed as an $N$-mode optical interferometer.
\item The term $\color{purple}H_{NG}$ corresponds to the difficult, nonlinear part of the calculation, and can be termed a controlled-squeezing term, i.e., qumodes $j\pm1$ being squeezed by a value determined by the field in qumode $j$. Such cubic gates can be implemented using alternating Gaussian two-mode squeezers (in blue) and non-Gaussian single-mode cubic phase gates (in red)~\cite{Kalajdzievski2019}

  \begin{gather}\label{eq:2cpg}
e^{i3\alpha t P_jQ_k^2} = \color{blue}e^{i2\alpha Q_jQ_k}\color{purple}e^{itP_j^3}\color{blue}e^{-i\alpha Q_jQ_k}\color{purple}e^{-itP_j^3}\color{blue}e^{-i2\alpha Q_jQ_k}\color{purple}e^{itP_j^3}\color{blue}e^{i\alpha Q_jQ_k}\color{purple}e^{-itP_j^3}e^{i\frac34\alpha^3tQ_j^3}.
\end{gather}
  \end{enumerate}

\begin{figure}[ht]
\centerline{\includegraphics[width=.5\textwidth]{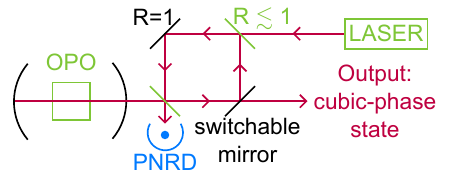}}
    \caption{Optical circuit for generating a cubic phase state. PNRD: photon-number-resolving detector. The components in green (OPO squeezing parameter, beamsplitter reflectivity, laser intensity) are controlled by the neural network, whose input data is from the components in blue (PNRD). The square delay loop also acts as an optical memory when the state is ready, in order to synchronize all such circuits, by switching all reflectivities to 1. The switchable mirror (e.g., Mach-Zehnder interferometer) allows the prepared state to continue to the rest of the circuit when ready.}
  \label{fig:cfg}
  \end{figure}

However, such a complicated gate decomposition does not necessarily need to be implemented directly. One of us (OP) recently discovered that non-Gaussian cubic phase gates such as $\exp(iP_j^3)$ can be directly and quasi-deterministically prepared by use of reinforcement learning (RL)~\cite{Anteneh2025}: a neural network can be trained to drive the quantum optical circuit of \fig{cfg}
and produce cubic-phase states $\exp(i\alpha Q^3)|0\rangle_p=\int ds \exp(i\alpha s^3)|s\rangle_q$, which can in turn implement single-mode cubic-phase gates $\exp(i\alpha Q^3)$ deterministically by (Gaussian) quantum teleportation~\cite{Gottesman2001}. Extension of this research to two-mode cubic phase gates such as the LHS of \eq{2cpg} is underway.

Another, fascinating avenue of research involves Trotterization of all components in a single OPO: note that \fig{cfg} describes the generation of a standalone cubic phase state $\exp(i\alpha Q^3)|0\rangle_p$ with full strength $\alpha$, whereas the quantum circuit Trotterizing \eq{HKdV} only requires strength $\alpha/p$ for $p$ Trotter steps. The idea will therefore be to extend our RL approach to combining all Trotterization components and yield the quantum evolution governed by \eq{HKdV}. A very preliminary, but initially much more feasible, experimental implementation could be to approximate the quantum evolution from \eq{HKdV} with $\exp(-i\tfrac\tau\hbar H_G)\exp(-i\tfrac\tau\hbar H_{NG})$, i.e., a 2$n$-mode interferometer fed by cubic phase states.

Once the whole KvN Hamiltonian of \eq{HKdV} is implemented, all we have to do is measure the amplitude quadratures $\{Q_j\}_{j=1,..., N}$ at the circuit's output. The measurement averages will yield the classical solutions $u_j=\avg{Q_j}$. Note that the initial state $u(t=0)$ can be encoded straightforwardly using displacement operations such as the one realized by the laser in \fig{cfg}.

\section{Concluding Remarks and Future Work\label{sec:con}}

The KvN formalism embeds classical dynamics in a Hilbert space using unitary evolution, enables simulation of classical systems with quantum algorithms, and facilitates use of quantum linear algebra to extract dynamical features of classical systems. One open question is that of reachable quantum advantage when simulating a classical system with a quantum machine. While the KvN formalism provides a systematic way to do this, one should remember that the mathematical problem space for an $N$-body classical system in 3 spatial dimensions, e.g.\ $N$ coupled oscillators, is the 6$N$-dimensional phase space of all position and momentum components, whereas the mathematical problem space for an $N$-qubit quantum system is the exponentially vaster $2^N$-dimensional Hilbert space of all quantum states. This remark constitutes the original case for quantum simulation made by Feynman~\cite{Feynman1982}. Beyond these dimensionality considerations, one can argue that a quantum machine could be beneficial when the classical computation is really hard, as in the case of a highly nonlinear Hamiltonian like the KdV one above. Other cases of interest might include nonintegrable classical and quantum systems. All these may provide interesting directions of investigation if the quantum machine can simulate faster than a classical machine can calculate, as quantum sampling and reconstructing expectation values on a full-scale quantum machine could be expected to be faster than classical numerics. Finally, one should also weigh the important advances of photonic quantum computing which can be leveraged in this context~\cite{Pfister2019}.

\bibliography{aiaa-q2}

\end{document}